\documentclass[twoside,twocolumn,english,aps,prb]{revtex4}
\usepackage[T1]{fontenc}
\usepackage[latin1]{inputenc}
\usepackage{amsmath}
\usepackage{graphicx}
\usepackage{amssymb}

\makeatletter

\providecommand{\LyX}{L\kern-.1667em\lower.25em\hbox{Y}\kern-.125emX\@}

\usepackage{babel}
\makeatother
\begin{document}

\title{Exact asymptotic form of the exchange interactions between shallow
centers in doped semiconductors}

\author{L.P. Gor'kov and P.L. Krotkov}

\affiliation{National High Magnetic Field Laboratory, Florida State University,
Tallahassee FL 32310,}

\affiliation{L.D. Landau Institute for Theoretical Physics, Russian Academy of
Sciences, 2 Kosygina st., 117334 Moscow Russia}

\date{June 09, 2003}

\begin{abstract}
The method developed in {[}L. P. Gor'kov and L. P. Pitaevskii, Sov.
Phys. Dokl. \textbf{8}, 788 (1964); C. Herring and M. Flicker, Phys.
Rev. \textbf{134}, A362 (1964){]} to calculate the asymptotic form
of exchange interactions between hydrogen atoms in the ground state
is extended to excited states. The approach is then applied to shallow
centers in semiconductors. The problem of the asymptotic dependence
of the exchange interactions in semiconductors is complicated by the
multiple degeneracy of the ground state of an impurity (donor or acceptor)
center in valley or band indices, crystalline anisotropy and strong
spin-orbital interactions, especially for acceptor centers in III-V
and II-VI groups semiconductors. Properties of two coupled centers
in the dilute limit can be accessed experimentally, and the knowledge
of the exact asymptotic expressions, in addition to being of fundamental
interest, must be very helpful for numerical calculations and for
interpolation of exchange forces in the case of intermediate concentrations.
Our main conclusion concerns the sign of the magnetic interaction
--- the ground state of a pair is always non-magnetic. Behavior of
the exchange interactions in applied magnetic fields is also discussed.

PACS number(s): 71.70.Gm, 71.55.Eq
\end{abstract}
\maketitle

\section{Introduction\label{sec:Introduction}}

\newcommand{\f}[1]{\mbox {\boldmath\(#1\)}}

\newcommand{\ff}[1]{\mbox{\scriptsize\boldmath\(#1\)}}

\newcommand{\ds}[1]{\displaystyle #1}

\newcommand{\ts}[1]{\textstyle #1}

\newcommand{\sgn}{\mathop{\mathrm{sgn}}\nolimits}

\newcommand{\const}{\mathop{\mathrm{const}}\nolimits}

Ferromagnetism in diluted III-V semiconductors has recently attracted
considerable attention for technologically appealing possibility of
integration of spin degrees of freedom into semiconductor devices.
In the most intensively studied compound of this group Ga$_{1-x}$Mn$_{x}$As
the ferromagnetic transition temperatures in thin annealed epitaxially
grown layers as high as $T_{C}>150$K for $x\sim 8$\% have been reported
\cite{ku03}.

The transition metal impurities in these materials substitute for
the atoms of the III group and act as both localized spins ($S=5/2$
for Mn due to half-filled $d$-shell) and as acceptors doping the
system with one hole. However, the system is highly compensated presumably
owing to double donor antisites (V group elements substituted for
III group elements) and metastable interstitial transition metal impurities.

A major not fully resolved issue is the mechanism governing ferromagnetism
in these compounds. Despite high disorder in these non-equilibrium
epitaxially grown films, RKKY interaction seems to be the most plausible
cause of Mn spin-spin interaction \cite{dietl01}. 

Meanwhile, we want to indicate that magnetic interactions in semiconductors
are not well understood yet even in a much simpler case of electrons
localized on non-magnetic impurities. Whereas two hydrogen atoms at
large distances are known to interact antiferromagnetically, the sign
of the interaction between donors/acceptors is not obvious because
of the band degeneracies as well as strong spin-orbital interaction
present. 

To the best of our knowledge, exchange interaction between donors
was considered only in Ref. \cite{andres81} in the Heitler-London
approximation which is known to be incorrect at large separations\cite{herring62,gorkov64,herring64}.
Besides, in Ref. \cite{andres81} an isotropic hydrogen-like model
was adopted in contrast to real semiconductors that have strong anisotropy
of conduction band valleys. In Section \ref{sec:Exchange-interaction-between}
below we compare our results for donors to those of Ref. \cite{andres81}.

The purpose of the present paper is to obtain \emph{asymptotically
exact} formulas for the exchange interaction between two shallow centers
at large separations. Although our results are pertinent to low impurity
concentrations, this information would be necessary for interpolating
results of numerical calculations to shorter inter-impurity separations
--- higher impurity concentrations.

The Heitler-London scheme, which considers exchange with the alien
atom as perturbation, is inapplicable \cite{herring62} to electron
exchange at large separations \textbf{$R$}, because the tunneling
takes place through a region where interactions with both atomic residues
and between the electrons themselves are comparable. As a matter of
fact, the Heitler-London expression for the exchange between two hydrogen
atoms incorrectly changes sign at large inter-atom separations. The
correct approach to the asymptotic exchange, that accurately accounted
for the interaction between the electrons and ionic residues during
the exchange process, was proposed in \cite{gorkov64,herring64}.
The magnitude of the exchange was expressed as an integral of the
probability current over the 5-dimensional median hyperplane separating
two regions of electron localization in their 6-dimensional coordinate
space.

Before proceeding to the corresponding problem in semiconductors we
list a number of extensions of the method \cite{gorkov64,herring64}
to other problems in atomic physics. In \cite{smirnov65} a generalization
was made for two atoms with non-equal binding energies of electrons
in $s$-states. It turned out, however, that in order for the method
to be applicable, the two energies should be close, which severely
restricted the applicable range of $R$ (see the discussion in \cite{chibisov88}).
In Ref. \cite{umanskii68} exchange integrals between degenerate $s$-
and $p$- states have been calculated starting from Wigner-Witmer
classification of the molecular states (see also \cite{nikitin84}).
Exchange integrals between higher orbital momentum $l$ states were
calculated in \cite{hadinger93}. In papers \cite{ponomarev99} the
method \cite{gorkov64,herring64} was applied to get interpolated
formulas down to small $R$ and to study two-dimensional electrons
localized on out-of-plane impurities. But to the best of our knowledge,
multiple degeneracy of donor or acceptor levels in real semiconductors
has never been addressed.

In this paper we extend the median-plane method to asymptotic exchange
between shallow impurities in semiconductors with localized carriers
in degenerate and non-spherically symmetric bands. We also study how
spin-orbital interaction and external magnetic field influence the
exchange.

As is well known, the basis of description of shallow impurity centers
in semiconductors is the effective mass method \cite{luttinger55,luttinger56}.
Shallow ceneters have small ionization energy compared to the forbidden
gap, and, consequently, the localization scale much larger than the
lattice constant. Hence for shallow centers the potential binding
carriers to the oppositely charged impurity ion may be deemed of approximately
Coulomb form: $V(\mathbf{r})=-e^{2}/\kappa r$, where $\kappa $ is
the permittivity. Following standard Kohn-Luttinger formalism \cite{luttinger55,luttinger56},
the wave function may then be expanded in many-electron Bloch functions
$\psi ^{(\alpha )}(\mathbf{r})$ at the extremum of the free band\begin{equation}
\psi ^{(\alpha )}(\mathbf{r})=\varphi ^{(\alpha )}(\mathbf{r})\psi ^{(\alpha )}(\mathbf{r}),\label{eq:wfrule}\end{equation}
 where $\alpha =1,\ldots ,\Gamma $, and $\Gamma $ is the band degeneracy
at the extremum, and the envelope function $\varphi ^{\alpha }(\mathbf{r})$
satisfies a Schr\"odinger equation\begin{equation}
E\varphi ^{\alpha }=-\ts \frac{1}{2}D_{ij}^{\alpha \beta }\partial _{i}\partial _{j}\varphi ^{\beta }+V(\mathbf{r})\varphi ^{\alpha }\label{eq:SchEq}\end{equation}
with the effective mass tensor $D_{ij}^{\alpha \beta }$ determined
by the crystal symmetry. 

We have generalized the approach \cite{gorkov64,herring64} to the
form of a secular equation, and in Section \ref{sec:Two-hydrogen-like-atoms}
of the paper we demonstrate the method on the example of two \emph{atoms}
with degenerate states. The atomic problem has much in common with
the case of two acceptors. In Section \ref{sec:Exchange-interaction-between}
we treat shallow donors in III-V (and also group IV) semiconductors
where the conduction band has several equivalent minima. Section \ref{sec:Asymmetric-exchange-interaction}
addresses the question of asymmetric exchange in III-V semiconductors.
Section \ref{sec:Degenerate-bands-(acceptors)} deals with acceptor
impurities localizing holes from degenerate valence bands. In Section
\ref{sec:Role-of-magnetic} the dependence of the exchange on the
applied magnetic field is studied. In the closing section we summarize
our results.

\section{Two hydrogen-like atoms --- secular equation approach\label{sec:Two-hydrogen-like-atoms}}

Consider two identical atoms A and B located at $-\mathbf{R}/2\equiv -R\hat{\f \zeta /2}$
and $\mathbf{R}/2$ respectively, each having one valent electron
outside of closed shells. Let an isolated atom have degenerate states
$\varphi _{\nu }(\mathbf{r})$ with binding energy $-\alpha ^{2}/2$.
Here $\nu $ are the quantum numbers describing the state. In central-field
approximation $\nu $ stands for a set of principal quantum number
$n$, orbital momentum $l$, its projection $m$ and spin projection
$s$. Henceforth we use atomic units, i.e. measure lengths in units
of the Bohr radius $a_{\mathrm{B}}=\hbar ^{2}/m_{e}e^{2}$, energies
in units of $2\mathrm{Ry}=m_{e}e^{4}/\hbar ^{2}$.

At large distances \begin{equation}
\varphi _{\nu }(\mathbf{r})\approx C_{\nu }Y_{l}^{m}(\hat{\mathbf{r}})r^{\frac{1}{\alpha }-1}e^{-\alpha r},\label{decay}\end{equation}
where $Y_{l}^{m}(\hat{\mathbf{r}})$ is a spherical function, and
$C_{\nu }$ is the asymptotic coefficient. 

The two-electron wave function $\psi ^{\alpha \beta }(\mathbf{r}_{1},\mathbf{r}_{2})$
in the absence of spin-orbital interaction can be factorized into
the product of spin $\chi ^{\alpha \beta }$ and coordinate $\psi (\mathbf{r}_{1},\mathbf{r}_{2})$
parts, where the former may be either a singlet $S=0$ or a triplet
$S=1$, and the latter is an eigenfunction of the Hamiltonian\begin{eqnarray}
\hat{\mathcal{H}} & = & -\ts \frac{1}{2}\f \partial _{1}^{2}-\ts \frac{1}{2}\f \partial _{2}^{2}-1/r_{1\mathrm{A}}-1/r_{2\mathrm{B}}\label{Ham0}\\
 & - & 1/r_{2\mathrm{A}}-1/r_{1\mathrm{B}}+1/r_{12}+1/R.\nonumber 
\end{eqnarray}
Here $\mathbf{S}=\mathbf{s}_{1}+\mathbf{s}_{2}$ is the total spin
and $\mathbf{r}_{1\mathrm{A}}=\mathbf{r}_{1}+\mathbf{R}/2$, $\mathbf{r}_{1\mathrm{B}}=\mathbf{r}_{1}-\mathbf{R}/2$,
etc.

In (\ref{Ham0}) the first row is the Hamiltonian of two isolated
atoms, while the terms in the second row result from bringing the
two atoms to a finite separation. These terms are taken as perturbation
in the Heitler-London approximation. The correct treatment of these
terms has to be accomplished in two stages.

First, if the possibility for the electrons to tunnel between the
two potential wells is neglected, this part of the Hamiltonian may
be expanded in multipole moments over inverse powers of $R$. Perturbation
theory calculations then yield van der Waals corrections to the energy
of each electron. We denote the energy of two electrons with van der
Waals interaction taken into account as $E_{0}$. Corresponding corrections
to the direct product $\varphi _{\nu _{\mathrm{A}}}(\mathbf{r}_{1})\varphi _{\nu _{\mathrm{B}}}(\mathbf{r}_{2})$
of isolated-atom wave functions retain electron localization. We denote
these perturbed wave functions when the first electron is localized
on ion A in the state $\nu _{\mathrm{A}}$ and the second --- on ion
B in the state $\nu _{\mathrm{B}}$ as $\psi _{\nu _{\mathrm{A}}\nu _{\mathrm{B}}}$.

Second, overlap of the wave functions of electrons (\ref{decay})
on the two atoms enables them to exchange position by tunneling. This
lifts the original degeneracy in energies of the isolated-atom states.
In contrast to van der Waals interactions, the resulting exchange
splitting, which exponentially decreases with distance, cannot be
calculated perturbatively. One can use symmetry reasoning to construct
the correct two-electron wave functions from the localized wave functions
$\psi _{\nu _{\mathrm{A}}\nu _{\mathrm{B}}}$ and interchanged $\hat{P}\psi _{\nu _{\mathrm{A}}\nu _{\mathrm{B}}}\equiv \psi _{\nu _{\mathrm{B}}\nu _{\mathrm{A}}}$.
Then the energy splitting is found by substituting those combinations
into Hamiltonian (\ref{Ham0}).

In the simplest case when atomic states are each only doubly degenerate
in spin projection $s$, the correct two-electron functions are symmetric
and antisymmetric combinations of the localized wave functions \cite{gorkov64}.
As total fermionic wave function must be antisymmetric in particle
permutation $\psi ^{\alpha \beta }(\mathbf{r}_{1},\mathbf{r}_{2})=-\psi ^{\beta \alpha }(\mathbf{r}_{2},\mathbf{r}_{1})$,
the singlet state corresponds to a symmetric coordinate wave function
and the triplet state --- to an antisymmetric one. Hence the energy
splitting between triplet and singlet states is given by\begin{equation}
E_{\mathrm{ex}}=-2J(R)(\mathbf{s}_{1}\mathbf{s}_{2}+1/4),\label{eq:Vex}\end{equation}
where $J(R)\propto e^{-2\alpha R}$ is found by substituting symmetric
and antisymmetric combinations into Hamiltonian (\ref{Ham0}) (see
below).

However, if degeneracy of atomic states is higher, the choice of correct
combination of two-electron wave functions becomes more complicated.
Higher orbital momentum $l$ states of two atoms are degenerate in
its projection $m$. In this case, the Wigner-Witmer classification
of molecular states in the absolute value of the total angular momentum
$\Lambda $ (and also parity for homo-atomic molecules) can be used
to identify the correct combinations of localized wave functions \cite{nikitin84}.
The combination satisfying the required symmetry constrains, however,
can sometimes be not unique. Or else, such a classification in cases
other than atomic may be complicated or absent.

We argue that one can simplify the procedure and find the correct
coefficients in the expansion \begin{equation}
\psi =\sum _{\nu _{\mathrm{A}}\nu _{\mathrm{B}}}\left(c_{\nu _{\mathrm{A}}\nu _{\mathrm{B}}}\psi _{\nu _{\mathrm{A}}\nu _{\mathrm{B}}}+c_{\nu _{\mathrm{A}}\nu _{\mathrm{B}}}^{(P)}\hat{P}\psi _{\nu _{\mathrm{A}}\nu _{\mathrm{B}}}\right)\label{expan}\end{equation}
as well as the energy level splitting in an approach similar to the
secular equation of perturbation theory of degenerate level. To this
end one substitutes (\ref{expan}) into two-electron wave equation
with the Hamiltonian (\ref{Ham0}) and uses $\hat{H}\psi _{\nu _{\mathrm{A}}\nu _{\mathrm{B}}}=E_{0}\psi _{\nu _{\mathrm{A}}\nu _{\mathrm{B}}}$.

In the next few paragraphs we derive an analogue for secular equation
in this approach. Let us denote the set of indices $\nu _{\mathrm{A}}\nu _{\mathrm{B}}$
by a single letter $n$. We start from the matrix continuity equation
$i\hbar \partial _{t}\psi _{i}^{*}\psi _{k}=-i\hbar \f \partial \mathbf{j}_{ik}$,
which is a direct consequence of the Schr\"odinger equation and where
the probability current density\begin{equation}
\mathbf{j}_{ik}=-\frac{i\hbar }{2m}\left(\psi _{i}^{*}\f \partial \psi _{k}-\psi _{k}\f \partial \psi _{i}^{*}\right).\label{eq:current}\end{equation}
Writing it for $\psi _{i}=\psi _{n}$ and $\psi _{k}=\psi $ from
(\ref{expan}) and integrating over the region $\zeta _{1}<\zeta _{2}$,
we obtain\begin{equation}
(E-E_{0})\int _{\zeta _{1}<\zeta _{2}}\psi _{n}^{*}\psi d^{6}\mathbf{r}_{1,2}=-i\hbar \int _{\zeta _{1}=\zeta _{2}}\mathbf{j}_{n,\psi }\mathbf{dS}.\label{eq:E}\end{equation}
Here $\mathbf{dS}=(\hat{\f \zeta _{1}},-\hat{\f \zeta _{2}})d\zeta d^{4}\f \rho _{1,2}$,
where $\f \rho _{1,2}$ are the transverse radii-vectors of the electrons,
is the element of the median 5-dimensional hyperplane $\zeta =\zeta _{1}=\zeta _{2}$
that bounds the integration region.

Functions $\psi _{n}$ describe electrons localized on different ions,
therefore the integral in the left hand side of Eq. (\ref{eq:E})
equals $c_{n}$ up to an exponentially small quantity which we neglect
as being of the same smallness as the energy splitting $E-E_{0}$
itself. From (\ref{eq:current}) we then obtain \begin{eqnarray}
(E & - & E_{0})c_{n}=\sum _{m}\int _{\zeta _{1}=\zeta _{2}}\bigl [\left(\psi _{m}\partial _{\zeta _{12}}\psi _{n}^{*}-\psi _{n}^{*}\partial _{\zeta _{12}}\psi _{m}\right)c_{m}\nonumber \\
 & + & \left(\hat{P}\psi _{m}\partial _{\zeta _{12}}\psi _{n}^{*}-\psi _{n}^{*}\partial _{\zeta _{12}}\hat{P}\psi _{m}\right)c_{m}^{(P)}\bigr ]d\zeta d^{4}\f \rho _{1,2},
\end{eqnarray}
where $\zeta =\frac{1}{2}(\zeta _{1}+\zeta _{2})$ and $\zeta _{12}=\zeta _{1}-\zeta _{2}$.

Differentiating the functions $\psi _{m}$ and $\hat{P}\psi _{m}$
in this expression one may retain only derivatives from the rapidly
decaying exponents $\psi _{m}\sim e^{-\alpha (R+\zeta _{12})}$, so
$\partial _{\zeta _{12}}\psi _{m}\approx -\alpha \psi _{m}$, and
$\partial _{\zeta _{12}}\hat{P}\psi _{m}\approx \alpha \hat{P}\psi _{m}$.
In this approximation the expression in the first parentheses vanishes
and we arrive at \begin{equation}
(E-E_{0})c_{n}=\sum _{m}J_{nm}c_{m}^{(P)},\label{eq:E-E0}\end{equation}
with the exchange integral matrix elements given by \begin{equation}
J_{nm}=-2\alpha \int _{\zeta _{1}=\zeta _{2}}\psi _{n}^{*}\hat{P}\psi _{m}d\zeta d^{4}\f \rho _{1,2}.\label{Vij2}\end{equation}

Likewise, starting from the permuted states $\hat{P}\psi _{n}$ we
integrate over the region $\zeta _{1}>\zeta _{2}$. Combining the
result with (\ref{eq:E-E0}), we get \begin{equation}
\left(\begin{array}{cc}
 0 & J_{nm}\\
 J_{nm} & 0\end{array}
\right)\left({c_{m}\atop c_{m}^{(P)}}\right)=(E-E_{0})\left({c_{n}\atop c_{n}^{(P)}}\right)\label{matrix}\end{equation}
(the Einstein convention on summation is implied). Since coordinate
wave functions of the triplet states are odd in electron interchange
($c_{n}=-c_{n}^{(P)}$) and those of the singlet states are even ($c_{n}=c_{n}^{(P)}$),
eigenvalue equation simplifies to\begin{equation}
J_{nm}c_{m}=\pm (E-E_{0})c_{n},\label{secular}\end{equation}
 where sign plus corresponds to singlet (even) solutions, sign minus
to triplet (odd) ones. We conclude that the splitting of originally
degenerate state $|\nu _{\mathrm{A}}\rangle \otimes |\nu _{\mathrm{B}}\rangle $
due to exchange must be found as eigenvalues of the exchange integral
matrix $J_{\nu _{\mathrm{A}}\nu _{\mathrm{B}},\nu '_{\mathrm{A}}\nu '_{\mathrm{B}}}$
and the coefficients in combinations (\ref{expan}) of localized wave
functions as corresponding eigenvectors.

The rest of the Section is devoted to the details of evaluation of
exchange integrals in atomic case, which will come useful for treatment
of acceptors below. To calculate $J_{\nu _{\mathrm{A}}\nu _{\mathrm{B}},\nu '_{\mathrm{A}}\nu '_{\mathrm{B}}}$
one needs to know the localized wave functions $\psi _{\nu _{\mathrm{A}}\nu _{\mathrm{B}}}$.
We are only interested in asymptotic dependence of exchange at large
$R$, hence we will find $\psi _{\nu _{\mathrm{A}}\nu _{\mathrm{B}}}$
in principal order in $R$. As in \cite{gorkov64,herring64}, we seek
$\psi _{\nu _{\mathrm{A}}\nu _{\mathrm{B}}}$ in the form \begin{equation}
\psi _{\nu _{\mathrm{A}}\nu _{\mathrm{B}}}(\mathbf{r}_{1},\mathbf{r}_{2})=\chi (\mathbf{r}_{1},\mathbf{r}_{2})\varphi _{\nu _{\mathrm{A}}}(\mathbf{r}_{1\mathrm{A}})\varphi _{\nu _{\mathrm{B}}}(\mathbf{r}_{2\mathrm{B}}),\label{eq:chi}\end{equation}
where $\chi $ is a smooth function varying on the scale of $R$.
Substituting (\ref{eq:chi}) into the Schr\"odinger equation $\hat{H}\psi _{\nu _{\mathrm{A}}\nu _{\mathrm{B}}}=E_{0}\psi _{\nu _{\mathrm{A}}\nu _{\mathrm{B}}}$,
where $E_{0}=-\alpha ^{2}+O(R^{-6})$ and keeping only terms of order
$R^{-1}$ (and thus neglecting the second derivatives of $\chi $),
we obtain\begin{equation}
\left[\alpha \left(\partial _{\zeta _{1}}-\partial _{\zeta _{2}}\right)-\frac{1}{R/2-\zeta _{1}}-\frac{1}{R/2+\zeta _{2}}+\frac{1}{R}+\frac{1}{r_{12}}\right]\chi =0.\label{eq:chieq}\end{equation}

Introducing the new variables $\zeta =\frac{1}{2}(\zeta _{1}+\zeta _{2})$
and $\zeta _{12}=\zeta _{1}-\zeta _{2}$, $\rho _{12}=|\f \rho _{1}-\f \rho _{2}|$,
in which $\partial _{\zeta _{1}}-\partial _{\zeta _{2}}=2\partial _{\zeta _{12}}$,
the solution of this equation is \begin{equation}
\chi =C(\zeta ,\rho _{12})\left(\frac{\sqrt{\sqrt{\zeta _{12}^{2}+\rho _{12}^{2}}-\zeta _{12}}}{(R-\zeta _{12}-2\zeta )(R-\zeta _{12}+2\zeta )}e^{-\zeta _{12}/2R}\right)^{\frac{1}{\alpha }},\end{equation}
where the integration constant $C(\zeta ,\rho _{12})$ must be determined
from the boundary condition $\chi \to 1$ when either $\mathbf{r}_{1}\to -\mathbf{R}/2$
or $\mathbf{r}_{2}\to \mathbf{R}/2$.

On the median plane $\zeta _{1}=\zeta _{2}=\zeta $ one thus obtains
\begin{equation}
\chi =\left(\frac{R}{R/2+|\zeta |}\sqrt{\frac{\rho _{12}}{R/2-|\zeta |}}e^{\frac{|\zeta |-R/2}{R}}\right)^{1/\alpha }.\label{chi}\end{equation}

We now substitute (\ref{eq:chi}) into the integrals in exchange elements
(\ref{Vij2}). Expanding the asymptotics (\ref{decay}): $e^{-\alpha r}\approx e^{-\alpha (|\zeta |+\rho ^{2}/2|\zeta |)}$,
the integrals over transverse coordinates $\f \rho _{1,2}$ acquire
the Gaussian form and converge rapidly on scale $\rho \sim \sqrt{R}$.
We thus need the asymptotics (\ref{decay}) only in the narrow cylinder
$\rho \sim \sqrt{R}\ll \zeta \sim R$, i.e. at small polar angles
$\theta $ from the internuclear axis $\hat{\f \zeta }$ chosen as
the quantization axis for orbital momentum. Since $Y_{l}^{m}(\theta ,\varphi )\approx p_{l}^{m}\theta ^{|m|}e^{im\varphi }/\sqrt{4\pi }$
when $\theta \ll 1$, where \begin{equation}
p_{l}^{m}=\frac{(-)^{\frac{m+|m|}{2}}}{2^{|m|}|m|!}\sqrt{(2l+1)\frac{(l+|m|)!}{(l-|m|)!}},\label{plm}\end{equation}
Eq. (\ref{decay}) may be rewritten as ($\theta \approx \rho /|\zeta |$)\begin{equation}
\varphi _{\nu }(\mathbf{r})\approx \frac{A_{\nu }}{\sqrt{4\pi }}\rho ^{|m|}|\zeta |^{\frac{1}{\alpha }-1-|m|}e^{-\alpha (|\zeta |+\rho ^{2}/2|\zeta |)}e^{im\varphi },\label{intrg}\end{equation}
 where $A_{\nu }=C_{\nu }p_{l}^{m}$ for atom A, and $A_{\nu }=(-1)^{l-m}C_{\nu }p_{l}^{m}$
for atom B. The sign multiplier for atom B originates from the change
of $\theta $ to $\pi -\theta $.

If all $m=0$ the exchange matrix elements $J$ may be evaluated with
the help of the substitution \begin{equation}
\f \rho =\ts \frac{1}{2}(\f \rho _{1}+\f \rho _{2}),\qquad \f \rho _{12}=\f \rho _{1}-\f \rho _{2},\label{eq:subst}\end{equation}
 which yields \cite{smirnov65} \begin{equation}
J_{\nu _{\mathrm{A}}\nu _{\mathrm{B}}\nu '_{\mathrm{A}}\nu '_{\mathrm{B}}}=-A_{\nu _{\mathrm{A}}}^{*}A_{\nu _{\mathrm{B}}}^{*}A_{\nu '_{\mathrm{A}}}A_{\nu '_{\mathrm{B}}}I(\alpha )R^{7/2\alpha -1}e^{-2\alpha R},\label{Vij4}\end{equation}
where \begin{equation}
I(\alpha )=\frac{\Gamma \left(1/2\alpha \right)}{2^{2+1/\alpha }\alpha ^{2+1/2\alpha }}\int _{0}^{1}q^{3/2\alpha }(2-q)^{1/2\alpha }e^{-q/\alpha }dq.\end{equation}
For non-degenerate $s$-states $p_{0}^{0}=1$(see (\ref{plm})), $A_{0}=C_{0}$
and the energies of singlet and triplet states are \begin{equation}
E_{s,t}=E_{0}\mp |C_{0}|^{4}I(\alpha )R^{7/2\alpha -1}e^{-2\alpha R}.\end{equation}

The asymptotic parameter $C_{0}$ of the atomic function (\ref{decay})
depends on the generally unknown form of the atomic potential $V(\mathbf{r})$
at small $r$. For a hydrogen atom in an $s$-state $C_{0}=2$, $\alpha =1$,
$I(1)\approx 0.511$ and we regain that the ground symmetric singlet
and excited antisymmetric triplet terms are separated by \cite{herring64}
$2\cdot 0.818R^{5/2}e^{-2R}$.

To find the hierarchy of exchange splittings in the general case,
one needs expressions for $J$ for other values of $m$ given in the
Appendix.

\section{Exchange interaction between shallow donors\label{sec:Exchange-interaction-between}}

In Ge, Si and some zinc blende semiconductors (GaP, AlSb) several
equivalent conduction band minima are located in equivalent points
of the Brillouin zone, turning one into each other under cubic symmetry
transformations. In Ge these are four $L$-points of intersection
of third order $\Lambda $-axes {[}111{]} with the Brillouin zone
boundary. In Si and III-V semiconductors minima lie on the three fourth
order $\Delta $-axes {[}100{]} at the $X$-points (III-V) on the
Brillouin zone boundary or close to them (Si). Electrons in the coaxial
pair of valleys in Si are indistinguishable, and such a pair may be
considered as one with doubled density of states. 

Isoenergy surfaces in the vicinity of a minimum numbered by the index
$u$ are ellipsoids of revolution around the symmetry axis $\hat{\mathbf{d}}_{u}$,
and the effective-mass tensor has uniaxial structure\begin{equation}
D_{uij}=m_{\bot }^{-1}(\delta _{ij}-\hat{d}_{ui}\hat{d}_{uj})+m_{\parallel }^{-1}\hat{d}_{ui}\hat{d}_{uj},\label{eq:valleyHk}\end{equation}
where $m_{\bot ,\parallel }$ are the transverse and longitudinal
effective masses of electrons and $\hat{d}_{ui}$ are the direction
cosines of the axis $\hat{\mathbf{d}}_{u}$ in a certain coordinate
system. In atomic units defined now as $a_{\mathrm{B}}=\kappa /m_{\bot }e^{2}$
and $2\mathrm{Ry}=m_{\bot }e^{4}/\kappa ^{2}$, i.e. corresponding
to the transverse mass: \begin{equation}
D_{uij}=(\delta _{ij}-\hat{d}_{ui}\hat{d}_{uj})+\gamma \hat{d}_{ui}\hat{d}_{uj},\label{degekinE}\end{equation}
where $\gamma =m_{\bot }/m_{\parallel }$.

\begin{figure}
\includegraphics[  width=0.90\columnwidth]{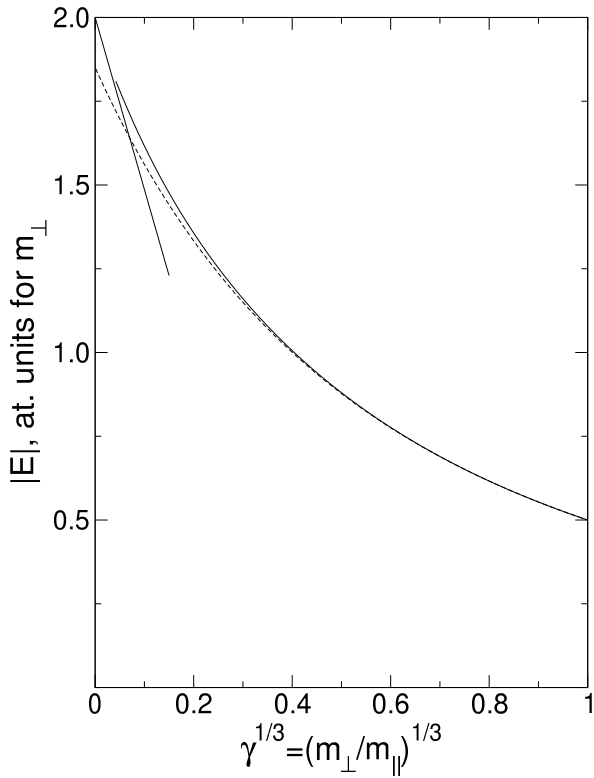}

\caption{Binding energy of electron in an elliptical valley with the transverse
to longitudinal mass ratio $\gamma $. Solid curve --- result of numerical
computations of the present paper. Dashed curve --- result of variational
fit with the trial function (\ref{eq:variational}). Solid straight
line --- approximate adiabatic solution (\ref{eq:adiabatic}) near
$\gamma =0$.\label{cap:Binding-energy-of}}
\end{figure}

Before proceeding to calculation of exchange interaction between elliptic
valleys we first remind several details about single-electron ground
state that will be important for the calculation. Ground state energy
spans the interval changing from $-2$ to $-\frac{1}{2}$ when $\gamma $
ranges from 0 to 1 (see Fig. \ref{cap:Binding-energy-of}). $\gamma =1$
corresponds to the spherical symmetric case with the ground state
energy $E_{t}=-\frac{1}{2}$ of a hydrogen-like center. For $\gamma <1$
ground state energy was calculated in \cite{kohn55} by variational
two-parameter fit with the trial function \begin{equation}
\varphi _{\mathrm{var}}=(\pi a_{\perp }^{2}a_{\parallel })^{-1/2}e^{-\sqrt{\rho ^{2}/a_{\perp }^{2}+z^{2}/a_{\parallel }^{2}}}\label{eq:variational}\end{equation}
(dashed curve in Fig. \ref{cap:Binding-energy-of}). 

Our results of numerical calculations by the method of differential
sweeping in an infinite interval with singularity \cite{abramov81}
are presented in Fig. \ref{cap:Binding-energy-of} as solid curve.
The wave function was sought in the form of the expansion in $L=40$
first Legendre polynomials in polar angle $\theta $ with respect
to the axis $\hat{\mathbf{d}}_{u}$. The cut-off at $L=40$ of the
indefinite basis cannot affect noticeably the accuracy of calculations
until the ratio of characteristic longitudinal and transverse length
scales $\gamma ^{1/2}$ becomes comparable or less than the minimal
scale $\sim L^{-1}$ of the truncated basis. For that reason we have
not plotted the numerical curve $E_{\parallel }(\gamma )$ down to
$\gamma \to 0$, where larger and larger basis would be required.
However, an approximate solution may be found in an adiabatic approach
\cite{kohn55}:\begin{equation}
E_{\parallel }\approx -2-4\alpha '_{0}\sqrt[3]{2\gamma },\label{eq:adiabatic}\end{equation}
 where $\alpha '_{0}\approx -1.02$ is the first root of the derivative
$\mathrm{Ai}'$ of the Airy function. This expression is shown in
Fig. \ref{cap:Binding-energy-of} as solid straight line.

At large distances from impurity the wave function of a localized
state with $E_{\parallel }=-\alpha ^{2}/2$ asymptotically falls off
as \begin{equation}
\varphi \approx A(\theta )\widetilde{r}^{q/\alpha -1}e^{-\alpha \widetilde{r}},\label{eq:anisdecay}\end{equation}
where $\widetilde{r}=|\widetilde{r_{i}}|$, $\widetilde{r_{i}}=D_{ij}^{-1/2}r_{j}$
and $D_{ij}^{-1/2}=(\delta _{ij}-\hat{d}_{i}\hat{d}_{j})+\gamma ^{-1/2}\hat{d}_{i}\hat{d}_{j}$.
We introduced \begin{equation}
q^{2}(\theta )\equiv D_{ij}^{-1}\hat{\zeta }_{i}\hat{\zeta }_{j}=\sin ^{2}\theta +\gamma ^{-1}\cos ^{2}\theta ,\label{eq:q(theta)}\end{equation}
where $\theta $ is the angle between $\hat{\f \zeta }$ and $\hat{\mathbf{d}}$. 

In real semiconductors several degenerate elliptic valleys with different
$\hat{\mathbf{d}}_{u}$ ($\hat{\mathbf{d}}_{u}$ along the rotation
axis of the ellipsoid) contribute, generally speaking, to the wave
function and hence to the exchange energy. Before addressing the complications
due to the valley degeneracy, consider first two donors in a hypothetic
tetragonal semiconductor with only one elliptic valley allowed by
symmetry. We assume again the two-electron wave function to have the
form (\ref{eq:chi}), where, however, the two single-electron functions
$\varphi _{\nu _{\mathrm{A}}}(\mathbf{r}_{1\mathrm{A}})$, $\varphi _{\nu _{\mathrm{B}}}(\mathbf{r}_{2\mathrm{B}})$
should now be found from solution of the Schr\"odinger equation for
an elliptic valley. Proceeding as above, from a bilinear kinetic energy
$D_{ij}\partial _{i}\partial _{j}$ we obtain instead of Eq. (\ref{eq:chieq})
\begin{eqnarray}
\Bigl [-\left(D_{1ij}\frac{\partial _{1i}\varphi _{\mathrm{A}}}{\varphi _{\mathrm{A}}}\partial _{1j}+D_{2ij}\frac{\partial _{2i}\varphi _{\mathrm{B}}}{\varphi _{\mathrm{B}}}\partial _{2j}\right) & - & \label{eq:50}\\
-\frac{1}{R/2-\zeta _{1}}-\frac{1}{R/2+\zeta _{2}}+\frac{1}{R}+\frac{1}{r_{12}}\Bigr ]\chi  & = & 0.\nonumber 
\end{eqnarray}

At large distances $\mathbf{r}\approx \zeta \hat{\f \zeta }$ along
the $\hat{\f \zeta }$-direction $\widetilde{r}\approx q(\theta )|\zeta |$,
where $q(\theta )$is from (\ref{eq:q(theta)}), and then from (\ref{eq:anisdecay})
obtain\begin{equation}
-\frac{D_{ij}\partial _{j}e^{-\alpha \widetilde{r}}}{e^{-\alpha \widetilde{r}}}=\alpha \frac{r_{i}}{\widetilde{r}}\approx \frac{\alpha }{q(\theta )}\widehat{\zeta }_{i}\sgn \zeta .\label{eq:Dijdj}\end{equation}

\begin{figure}
\includegraphics[  width=0.70\columnwidth]{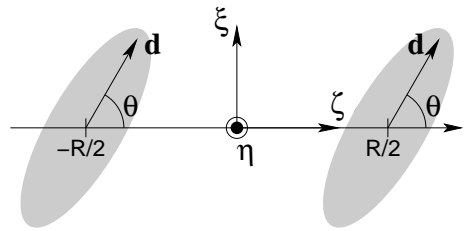}

\caption{\label{cap:tetragonal}Frame ($\hat{\f \xi },\hat{\f \eta },\hat{\f \zeta }$)
for calculation of exchange between two elliptic valleys with the
same $\hat{\mathbf{d}}$. The $\hat{\f \zeta }$-axis connects the
two centers.}
\end{figure}

In the coordinate system chosen the angles $\theta $ are equal for
the two impurity centers (see Fig. \ref{cap:tetragonal}). Hence (\ref{eq:50})
reduces to (\ref{eq:chieq}) with $\alpha \to \alpha /q(\theta )$.
This substitution also gives the new expression for the matrix elements
$J$ to replace (\ref{Vij2}). Also the two-electron wave functions
(\ref{eq:chi}) in expression (\ref{Vij2}) now contain single-electron
functions corresponding to elliptic valleys.

To calculate $J$ we now follow the scheme presented in Section \ref{sec:Two-hydrogen-like-atoms}
for atoms. To expand the exponents in the single-electron wave functions
we supplement the inter-impurity axis $\hat{\f \zeta }$ with perpendicular
axes $\hat{\f \xi }$ and $\hat{\f \eta }$ so that $\hat{\mathbf{d}}=\hat{\f \zeta }\cos \theta +\hat{\f \xi }\sin \theta $
and expand $\widetilde{r}$ in the small ratios $\xi /\zeta $ and
$\eta /\zeta $ up to second order \begin{equation}
\widetilde{r}\approx q|\zeta |\left[1+\frac{(\ts \frac{1}{\gamma }-1)\sin \theta \cos \theta }{q^{2}}\frac{\xi }{\zeta }+\frac{\eta ^{2}}{2q^{2}\zeta ^{2}}+\frac{\xi ^{2}}{2\gamma q^{4}\zeta ^{2}}\right].\label{eq:44}\end{equation}
Applying this expansion to the sum $\widetilde{r}_{1\mathrm{A}}+\widetilde{r}_{1\mathrm{B}}$
of the exponents we substitute $\zeta $ with $\zeta _{\pm }=\zeta \pm R/2$.
Since for the electrons located between the ions $-R/2<\zeta <R/2$:
$|\zeta _{\pm }|=\pm \zeta _{\pm }$, terms linear in $\xi $ in (\ref{eq:44})
cancel each other. 

Hence like in atomic case we are left with the Gaussian integrals
over transverse coordinates $\f \rho _{1}=(\xi _{1},\eta _{1})$ and
$\f \rho _{2}$ that decouple from the integral over $\zeta $. By
the substitution (\ref{eq:subst}) the Gaussian integrals reduce to
a single integral over the polar angle $\varphi _{12}$ of $\f \rho _{12}$:\begin{equation}
S(\theta )=\frac{1}{2\pi }\int _{0}^{2\pi }\frac{d\varphi }{\left[\sin ^{2}\varphi +\cos ^{2}\varphi /\gamma q^{2}\right]^{1+\frac{q}{2\alpha }}}.\end{equation}
In the spherically symmetric case $\gamma =q=1$ the integral $S(\theta )=1$. 

Calculating matrix element (\ref{Vij2}) gives eventually\begin{equation}
J=-(4\pi )^{2}\sqrt{\gamma }A^{4}(\theta )S(\theta )I(\alpha ,q)R^{7q/2\alpha -1}e^{-2\alpha qR},\label{eq:blinkakzamanalo}\end{equation}
where \begin{equation}
I(\alpha ,q)=\frac{\Gamma \left(q/2\alpha \right)q^{3+q/2\alpha }}{2^{2+q/\alpha }\alpha ^{2+q/2\alpha }}\int _{0}^{1}t^{3/2\alpha }(2-t)^{1/2\alpha }e^{-qt/\alpha }dt.\label{eq:35}\end{equation}
and the integration variable in (\ref{eq:35}) $t=1-2\zeta /R$. For
two spherically symmetric ($\gamma =1$) atoms with angular momentum
projection $m=0$ the asymptotic coefficient $A(\theta )=A/\sqrt{4\pi }$
(see (\ref{intrg})) and Eq. (\ref{eq:blinkakzamanalo}) reduces to
(\ref{Vij4}) as it should. 

\begin{figure}
\includegraphics[  width=0.99\columnwidth]{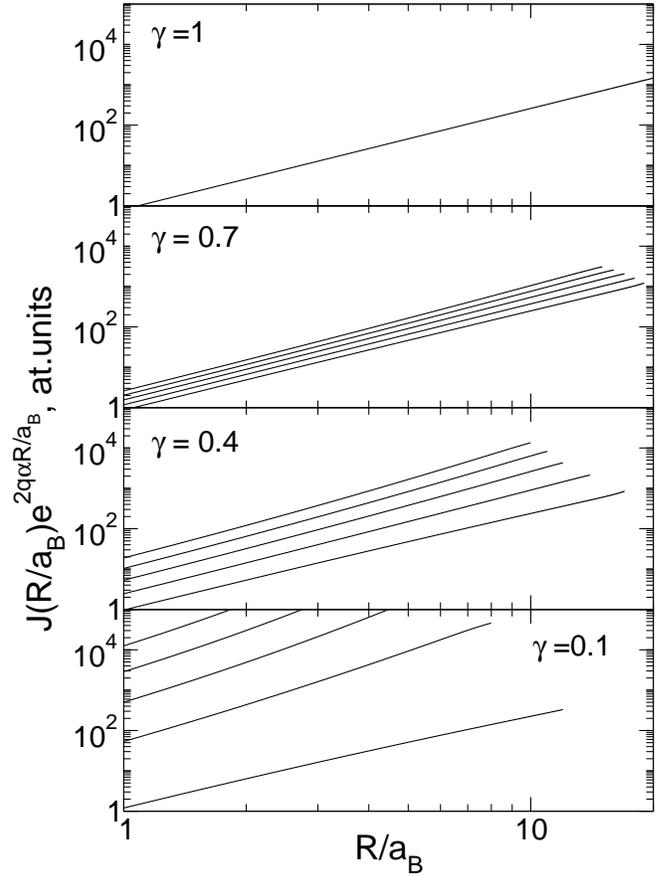}

\caption{Preexponents of exchange integrals between ground states of electrons
localized in elliptic valleys vs. donor center separation for four
values of $\gamma =0.1$, 0.4, 0.7, and 1. For each value of $\gamma $
five curves are plotted for different angles between the preferred
direction of the elliptic valley and the inter-center axis: $\theta =0$,
$\pi /6$, $\pi /4$, $\pi /3$, $\pi /2$ (from top to bottom). In
the spherical symmetric case $\gamma =1$ exchange does not depend
on $\theta $.\label{cap:Preexponents-of-exchange}}
\end{figure}

In addition to finding the binding energy we have also carried out
numerical calculation of single-electron wave functions to obtain
asymptotics (\ref{eq:anisdecay}) and exchange integrals (\ref{eq:blinkakzamanalo})
using the differential sweeping method \cite{abramov81}. The preexponents
of the exchange integrals, $J(R)e^{2\alpha qR}$, are presented in
Fig. \ref{cap:Preexponents-of-exchange} for $\gamma =0.1$, 0.4,
0.7, and 1, and for $\theta =0$, $\pi /6$, $\pi /4$, $\pi /3$,
and $\pi /2$ (from top to bottom) for each $\gamma $. In the interval
of inter-center separations $1<R<10$ accessible for numerical evaluation
the dependence of preexponents on $R$ is approximately linear in
double logarithmic scale with the slope agreeing well with $7q/2\alpha -1$
which follows from (\ref{eq:blinkakzamanalo}).

Returning now to realistic semiconductors, recall that degenerate
valleys have different anisotropic dependence for the decay of the
wave function of localized electron. For an arbitrary orientation
of the axis connecting two donor centers the main contribution to
exchange interaction comes from the wave function components corresponding
to the pair of valleys with the slowest decay, contributions of the
other pairs being exponentially suppressed. In other words, among
all the elements \begin{equation}
\langle uv|J|u'v'\rangle \sim e^{-\alpha [q(\theta _{u})+q(\theta _{v})+q(\theta _{u'})+q(\theta _{v'})]R/2},\label{eq:JNK}\end{equation}
 we can, with the accuracy up to exponentially small terms, keep only
the elements between such states $u$ that the exponent in (\ref{eq:JNK})
is minimal: $\langle uu|J|uu\rangle \sim e^{-2\alpha q(\theta _{u})R}$.
Therefore, for any arrangement of impurity centers except for high
symmetry directions the electron exchange is predominantly caused
by the overlap of the wave function components from valleys with $\hat{\mathbf{d}}_{\mathrm{A}}=\hat{\mathbf{d}}_{\mathrm{B}}$,
which means that the term splitting is correctly described by Eq.
(\ref{eq:blinkakzamanalo}). 

The relative position of the two donors being changed, crossover between
the pairs of valleys contributing the most into the exchange matrix
element (\ref{eq:JNK}) occurs %
\footnote{Along the directions very close to some symmetry directions (the four
axes of the third order in Si, and the three fourth-order axes in
Ge), it is necessary to take into account several valleys simultaneously.
Because the shift between exponential dependencies takes place in
a very small solid angle around the axis, we will not dwell on such
situation.%
}. 

These considerations are not changed if one takes into account the
lifting of the inter-valley degeneracy of a single center by the short-range
part of the impurity potential (``central-cell correction''). The
ground state in this case is the non-degenerate symmetric combination
of the functions (\ref{eq:wfrule})\cite{kohn55}\begin{equation}
\psi _{g}(\mathbf{r})=\frac{1}{\sqrt{\Gamma }}\sum _{u=1}^{\Gamma }\varphi _{u}(\mathbf{r})\psi _{u}(\mathbf{r}),\label{eq:gerade}\end{equation}
where $\psi _{u}(\mathbf{r})\propto e^{i\mathbf{k}_{u}\mathbf{r}}$
are the Bloch functions at the minimum of the conduction band located
at $\mathbf{k}=\mathbf{k}_{u}$, and $\Gamma $ is the number of the
participating valleys (6 in Si, 4 in Ge, and 3 in III-V compounds).

Considering the Bloch functions oscillate rapidly on atomic scale
much smaller than the effective Bohr radius, exchange interactions
between the states (\ref{eq:gerade}) may be seen to be\cite{andres81}\begin{equation}
J_{g}\approx \frac{1}{\Gamma ^{2}}\sum _{u,v=1}^{\Gamma }\langle uv|J|uv\rangle e^{i(\mathbf{k}_{v}-\mathbf{k}_{u})\mathbf{R}}.\end{equation}
If we average over the rapidly oscillating exponential terms, we get
\begin{equation}
J_{g}\approx \frac{1}{\Gamma ^{2}}\sum _{u=1}^{\Gamma }\langle uu|J|uu\rangle ,\end{equation}
which shows that the exchange between the symmetric ground states
of multi-valley donors for any arrangement except along high symmetry
directions is described by Eq. (\ref{eq:blinkakzamanalo}) with the
asymptotic parameter $A(\theta )\to A(\theta )/\sqrt{\Gamma }$.

Concluding this Section, we discuss briefly the previous study of
exchange interaction between donors \cite{andres81}. In the cited
paper, the anisotropy of the exponential dependence of exchange interaction
for elliptic valleys was completely ignored by substituting the expression
for isotropic atomic exchange. Whereas, e.g., for Si $\gamma =m_{\bot }/m_{\parallel }\approx 0.21$,
and hence $q(\theta )$ (\ref{eq:q(theta)}) in the exponent of the
exchange interaction $\propto e^{2\alpha q(\theta )R}$ changes from
$\gamma ^{-1/2}\approx 2.2$ for $\theta =0$ to $0$ for $\theta =\pi /2$.
Besides, in Ref. \cite{andres81} the Heitler-London approximation
was used, the approach proved incorrect in calculating the preexponential
factors at large distances. On the other hand, we obtained exact asymptotic
expressions for the dependence of the exchange interaction on the
valley parameter $\gamma $ and the mutual orientation of donors $\theta $.

\section{Asymmetric exchange interaction (donors)\label{sec:Asymmetric-exchange-interaction}}

In semiconductors without center of inversion symmetry, such as III-V
or II-VI compounds, spin-orbital interaction for donors acquires the
form \cite{dresselhaus55,DP}\begin{equation}
\hat{\mathcal{H}}_{\mathrm{so}}=\mathbf{h}(-i\f \partial )\hat{\mathbf{s}}.\label{eq:so0}\end{equation}
Wave function of a single electron cannot be represented as a product
of spin and coordinate components any more. An approach with the expansion
(\ref{expan}) is nevertheless applicable after suitable adjustments
\cite{gorkov03}.

In the presence of spin-orbital Hamiltonian (\ref{eq:so0}) spin projection
no longer commutes with the total Hamiltonian, but the two-fold (Kramers)
degeneracy of the ground state remains. Index numbering this degeneracy
may be included into the full set $\nu $ of indices describing a
single-electron state. The permutation operator $\hat{P}$ in (\ref{expan})
now interchanges both spin and coordinate variables of the electrons.

Correspondingly, the provision for the total wave function $\psi $
(\ref{expan}) to be antisymmetric in electron permutation: $\psi =-\hat{P\psi }$,
leads to $c_{\nu _{\mathrm{A}}\nu _{\mathrm{B}}}=-c_{\nu _{\mathrm{A}}\nu _{\mathrm{B}}}^{(P)}$.

For simplicity we will restrict ourselves to donors in II-VI and III-V
semiconductors where besides being a Kramers doublet single-electron
state on a donor is not otherwise degenerate. Then $\nu $ will stand
for the Kramers index.

In bulk zinc-blende semiconductors \cite{dresselhaus55,DP}\begin{equation}
h_{x}(\mathbf{k})=\frac{\alpha _{\mathrm{so}}}{\sqrt{2E_{g}}}k_{x}(k_{y}^{2}-k_{z}^{2})\label{eq:so}\end{equation}
 and $h_{y}(\mathbf{k})$ and $h_{z}(\mathbf{k})$ are obtained by
permutation in (\ref{eq:so}). Here $E_{g}$ is the band gap in atomic
units and $\alpha _{\mathrm{so}}$ is a phenomenological parameter.
For GaAs $E_{g}\approx 1.43$ eV (in atomic units this yields $E_{g}\approx 113$
under the radical sign in (\ref{eq:so})) and $\alpha _{\mathrm{so}}\approx 0.07$. 

For small spin-orbital interaction the spinor single-electron wave
functions are\begin{equation}
\varphi _{\nu }^{\beta }(\mathbf{r})\approx \varphi _{0}(r)(e^{i\alpha ^{2}\mathbf{h}(\hat{\mathbf{r}})r\ff \sigma /2})^{\beta \nu },\label{eq:phase}\end{equation}
where $\varphi _{0}(r)=e^{-r}/\sqrt{\pi }$ is the hydrogenic ground
state wavefunction and $\hat{\mathbf{r}}=\mathbf{r}/r$. The phase
factor turns into unity when $r\to 0$. Thus the Kramers index takes
on the meaning of the spin projection of an electron on the center
$\nu =\uparrow ,\downarrow $.

Two-electron wave functions have now to be sought as solutions of
the Hamiltonian \begin{eqnarray}
\hat{\mathcal{H}} & = & -\ts \frac{1}{2}\f \partial _{1}^{2}-\ts \frac{1}{2}\f \partial _{2}^{2}+\hat{\mathcal{H}}_{\mathrm{so}1}+\hat{\mathcal{H}}_{\mathrm{so}2}-1/r_{1\mathrm{A}}-1/r_{2\mathrm{B}}\label{Ham}\nonumber \\
 & - & 1/r_{2\mathrm{A}}-1/r_{1\mathrm{B}}+1/r_{12}+1/R.\label{eq:HamSO}
\end{eqnarray}
in the form (\ref{chi})\begin{equation}
\psi _{\nu _{\mathrm{A}}\nu _{\mathrm{B}}}^{\alpha \beta }(\mathbf{r}_{1},\mathbf{r}_{2})=\chi ^{\alpha \alpha ',\beta \beta '}(\mathbf{r}_{1},\mathbf{r}_{2})\varphi _{\nu _{\mathrm{A}}}^{\alpha '}(\mathbf{r}_{1\mathrm{A}})\varphi _{\nu _{\mathrm{B}}}^{\beta '}(\mathbf{r}_{2\mathrm{B}}),\label{eq:AB}\end{equation}
with $\chi ^{\alpha \alpha ',\beta \beta '}$converting to $\delta ^{\alpha \alpha '}\delta ^{\beta \beta '}$
when either $\mathbf{r}_{1\mathrm{A}}\to 0$ or $\mathbf{r}_{2\mathrm{B}}\to 0$.
The principal terms in spin-orbital interaction are included into
definition (\ref{eq:phase}), and the equation for $\chi ^{\alpha \alpha ',\beta \beta '}$
turns out to coincide with (\ref{eq:chieq}). Therefore $\chi ^{\alpha \alpha ',\beta \beta '}=\chi \delta ^{\alpha \alpha '}\delta ^{\beta \beta '}$
with $\chi $ from (\ref{eq:chi}).

Similarly to (\ref{eq:E-E0}) we get \begin{equation}
(E-E_{0})c_{\nu _{\mathrm{A}}\nu _{\mathrm{B}}}=\sum _{\nu '_{\mathrm{A}}\nu '_{\mathrm{B}}}J_{\nu _{\mathrm{A}}\nu _{\mathrm{B}}\nu '_{\mathrm{A}}\nu '_{\mathrm{B}}}c_{\nu '_{\mathrm{A}}\nu '_{\mathrm{B}}}^{(P)},\end{equation}
where \begin{eqnarray}
J_{\nu _{\mathrm{A}}\nu _{\mathrm{B}}\nu '_{\mathrm{A}}\nu '_{\mathrm{B}}} & = & -2\int _{\zeta _{1}=\zeta _{2}}\psi _{\nu _{\mathrm{A}}\nu _{\mathrm{B}}}^{*}\psi _{\nu '_{\mathrm{B}}\nu '_{\mathrm{A}}}d\zeta d^{4}\f \rho _{1,2}\nonumber \\
 & = & J(R)(e^{-i\ff \gamma \ff \sigma /2})_{\nu _{\mathrm{A}}\nu '_{\mathrm{B}}}(e^{i\ff \gamma \ff \sigma /2})_{\nu _{\mathrm{B}}\nu '_{\mathrm{A}}}.
\end{eqnarray}
The exchange integral $J(R)=-0.818R^{5/2}e^{-2R}$ \cite{gorkov64,herring64}
and \begin{equation}
\f \gamma =\mathbf{h}(\hat{\f \zeta })R..\label{eq:gamma}\end{equation}
Absolute value $|\f \gamma |/2$ plays the role of angle of rotation
of electron spin around the axis $\hat{\f \gamma }$ when it tunnels
between each of the centers A and B and the median plane. Using \begin{equation}
c_{\nu _{\mathrm{A}}\nu _{\mathrm{B}}}^{(P)}=-c_{\nu _{\mathrm{A}}\nu _{\mathrm{B}}}\equiv -\ts \frac{1}{2}(\f \sigma _{\nu _{\mathrm{A}}\nu '_{\mathrm{A}}}\f \sigma _{\nu _{\mathrm{B}}\nu '_{\mathrm{B}}}+\delta _{\nu _{\mathrm{A}}\nu '_{\mathrm{A}}}\delta _{\nu _{\mathrm{B}}\nu '_{\mathrm{B}}})c_{\nu '_{\mathrm{B}}\nu '_{\mathrm{A}}}.\label{eq:13}\end{equation}
 we finally arrive at the equation $(E-E_{0})c=\hat{\mathcal{H}}_{\mathrm{ex}}c$
for the coefficients $c_{\nu _{\mathrm{A}}\nu _{\mathrm{B}}}$, with
the Hamiltonian \cite{gorkov03}\begin{eqnarray}
\hat{\mathcal{H}}_{\mathrm{ex}} & = & -\ts \frac{1}{2}Je^{-i\ff \gamma (\ff \sigma _{1}-\ff \sigma _{2})/2}\left(\hat{\f \sigma }_{1}\hat{\f \sigma }_{2}+1\right)\label{eq:ee1}\\
 & \equiv  & -\ts \frac{1}{2}J\left[\hat{\f \sigma }_{1}\overleftrightarrow{R}(\f \gamma )\hat{\f \sigma }_{2}+1\right]\nonumber \\
 & \approx  & -\ts \frac{1}{2}J\left[\hat{\f \sigma }_{1}\hat{\f \sigma }_{2}+1-\f \gamma \left(\hat{\f \sigma }_{1}\times \hat{\f \sigma }_{2}\right)\right],\nonumber 
\end{eqnarray}
where the Pauli matrices $\hat{\f \sigma }_{1}$ and $\hat{\f \sigma }_{2}$
act in the space of the Kramers indices, and $\overleftrightarrow{R}(\f \gamma )$
is the three-dimensional rotation matrix on the angle $\gamma $ around
the axis $\hat{\f \gamma }$.

In the absence of spin-orbital interaction (\ref{eq:so0}) the angle
$|\f \gamma |=0$ and we return to the exchange Hamiltonian (\ref{eq:Vex})
as expected.

In this Section we have considered the expansion of spin-orbital interaction
(\ref{eq:so0}) in bulk III-V semiconductors, where (\ref{eq:so0})
begins with the third power in momentum. Analogously, one can study
spin-orbital interaction linear in momentum which appears in reduced
symmetry or dimensionality (in films or at the surface). All the derivation
including the expression (\ref{eq:gamma}) for angle $\f \gamma $
remains intact. The case of linear spin-orbital interaction has also
been studied differently in Ref. \cite{kavokin03}.

\section{Degenerate bands (acceptors)\label{sec:Degenerate-bands-(acceptors)}}

As we will see later, calculation of the exchange interaction between
two shallow acceptor centers has much in common with the problem of
two hydrogen atoms each in a state with non-zero momentum discussed
in Section \ref{sec:Two-hydrogen-like-atoms} and in the Appendix.
However, the analogy is not complete, the calculations become laborious
and we have not performed them fully. In what follows we only determine
the \emph{sign} of the exchange interaction. More precisely, it is
shown that the magnetic moment in the ground state of a pair of acceptors
equals zero.

We begin this Section by first reviewing the properties of a single
acceptor center.

Without taking spin-orbital interaction into account the valent band
of typical semiconductors (Ge, Si, III-V) would be six-fold degenerate
at the point $\mathbf{k}=0$. Spin-orbital interaction splits the
valent band into a two- and four-fold degenerate bands separated by
a gap. If the spin-orbital splitting is large with respect to the
binding energy of an acceptor, the two-fold degenerate band can be
neglected. For $\mathbf{k}\ne 0$ the four-fold degeneracy of the
upper-lying band is lifted and two two-fold bands of light and heavy
holes are formed. Their energy spectrum $\epsilon (\mathbf{k})$ quadratic
in $k$ is known to depend on the direction of the quasimomentum $\hat{\mathbf{k}}$
through three invariants of the $O_{h}$ point symmetry group so that
surfaces $\epsilon (\mathbf{k})=\const $ are slightly warped spheres.

As is known, the spectrum in the vicinity of $\mathbf{k}=0$ is determined
by three Luttinger parameters $\gamma _{1}$, $\gamma _{2}$ and $\gamma _{3}$.
In practice, two of them are close $\gamma _{2}\approx \gamma _{3}=\overline{\gamma }$
and the dispersion relation is simply $\epsilon =k^{2}/2m_{l,h}$,
where $m_{l,h}=m_{0}/(\gamma _{1}\pm \overline{\gamma })$ are the
effective masses of the light and heavy holes. In this so called spherical
approximation the Schr\"odinger equation for the four-component envelope
hole wave function $\varphi ^{\alpha }$ is given by (\ref{eq:SchEq})
with the effective mass tensor\begin{equation}
D_{ij}=\frac{1}{m_{0}}\left[\gamma _{1}\delta _{ij}-\overline{\gamma }\left(\{J_{i}J_{j}\}-\ts \frac{1}{3}\mathbf{J}^{2}\delta _{ij}\right)\right],\label{eq:semidij}\end{equation}
where $\mathbf{J}$ is the 4$\times $4 matrix of the pseudospin operator
$J=\frac{3}{2}$ and $\{J_{i}J_{j}\}=\frac{1}{2}(J_{i}J_{j}+J_{j}J_{i})$.

It is convenient to rewrite (\ref{eq:semidij}) using atomic units
$a_{\mathrm{B}}=\kappa /m_{h}e^{2}$ and $2\mathrm{Ry}=1/m_{h}a_{\mathrm{B}}^{2}$:
\begin{equation}
D_{ij}=\frac{1}{1-\mu }\left[\delta _{ij}-\mu \left(\{J_{i}J_{j}\}-\ts \frac{1}{3}\mathbf{J}^{2}\delta _{ij}\right)\right].\label{kinE}\end{equation}
Here \begin{equation}
\mu =\frac{\overline{\gamma }}{\gamma _{1}}=\frac{m_{h}-m_{l}}{m_{h}+m_{l}}.\end{equation}

In the limit of equal masses $m_{h}$ and $m_{l}$ the four components
of $\varphi ^{\alpha }$ in (\ref{eq:SchEq}) decouple, each satisfying
the Schr\"odinger equation for a hydrogen atom with the ground state
energy $E_{h}=-1/2$ for orbital momentum $L=0$. In the general case
$m_{h}>m_{l}$ acceptor states may be classified \cite{lipari70,gel'mont71,baldereschi73}
by the total angular momentum $\mathbf{F}=\mathbf{J}+\mathbf{L}$
and its projection $m_{F}$:\begin{equation}
|Fm_{F}\rangle =\sum _{L}\sum _{m_{J},m_{L}}\langle m_{J}m_{L}|Fm_{F}\rangle |Jm_{J}\rangle |Lm_{L}\rangle ,\label{eq:FmF}\end{equation}
where $\langle m_{J}m_{L}|Fm_{F}\rangle $ are the Clebsh-Gordan coefficients
and $m_{J}+m_{L}=m_{F}$, $|Lm_{L}\rangle =R_{L}(r)Y_{L}^{m_{L}}(\hat{\mathbf{r}})$,
$Y_{L}^{m_{L}}(\hat{\mathbf{r}})$ are the spherical harmonics and
$|Jm_{J}\rangle $ is a column in the pseudospin index with one non-zero
component at $m_{J}$. The summation over $L$ for a given $F$ runs
over all such values that $|L-J|\le F\le L+J$. 

Substituting (\ref{eq:FmF}) into (\ref{eq:SchEq}) with (\ref{kinE})
results in the Sturm-Liouville problem on the energy $E_{h}$ and
the unknown radial functions $R_{L}(r)$. The ground state turns out
to be four-fold degenerate with $F=\frac{3}{2}$ (It is, in fact,
the only state with possible $L=0$, hence it becomes the correct
ground state in the limit $m_{h}=m_{l}$). In this state only the
radial functions $R_{L}(r)$ with $L=0$ and 2 turn out to be coupled:\begin{eqnarray}
\left(\begin{array}{cc}
 \partial _{r}^{2}+\frac{2}{r}\partial _{r} & -\mu \left(\partial _{r}^{2}+\frac{5}{r}\partial _{r}+\frac{3}{r^{2}}\right)\\
 -\mu \left(\partial _{r}^{2}-\frac{1}{r}\partial _{r}\right) & \partial _{r}^{2}+\frac{2}{r}\partial _{r}-\frac{6}{r^{2}}\end{array}
\right)\left({R_{0}\atop R_{2}}\right) &  & \label{eq:R0R2}\\
+2(1-\mu )\left(E_{h}+\ts \frac{1}{r}\right)\left({R_{0}\atop R_{2}}\right) & = & 0\nonumber 
\end{eqnarray}

Numerical solution \cite{gel'mont71} of this equation shows that
$E_{h}$ changes from $-\frac{1}{2}$ at $\mu =0$ ($m_{h}=m_{l}$)
to $\approx -\frac{2}{9}$ in the limit $\mu \to 1$ ($m_{h}\gg m_{l}$).
Thus the binding energy is determined by the scale of the heavy mass.
At large $r$ (\ref{eq:R0R2}) yields \begin{eqnarray}
(1-\mu )(\partial _{r}^{2}+2E_{h})(R_{0}+R_{2}) & = & 0,\\
(\partial _{r}^{2}+2E_{h}m_{l}/m_{h})(R_{0}-R_{2}) & = & 0,\label{R0-R2}
\end{eqnarray}
which shows that for $m_{h}\gg m_{l}$ the sum of the two functions
falls off on the scale $1/\sqrt{2|E_{h}|}$ corresponding to the heavy
holes, i.e. much faster than their difference, falling off on the
scale determined by the light holes. Then at large $r$ \cite{shklovskii84}
$R_{0}\approx -R_{2}\propto e^{-\alpha _{l}r}$ with \begin{equation}
\alpha _{l}=\sqrt{2|E_{h}|m_{l}/m_{h}}.\end{equation}

The four degenerate wave functions of the ground state (\ref{eq:FmF})
in the asymptotic region close to the axis $\hat{\f \zeta }$ connecting
the impurities (chosen as the quantization axis for the orbital momentum
$\mathbf{L}$) are\begin{eqnarray}
\varphi _{\frac{3}{2}}(\mathbf{r}) & = & -R_{0}(r)\sqrt{\frac{3}{4\pi }}\theta e^{i\varphi }|\ts \frac{1}{2}\rangle ,\nonumber \\
\varphi _{\frac{1}{2}}(\mathbf{r}) & = & \frac{R_{0}(r)}{\sqrt{\pi }}\left(|\ts \frac{1}{2}\rangle -\frac{\sqrt{3}}{2}\theta e^{-i\varphi }|\ts \frac{3}{2}\rangle \right),\label{eq:phi}\\
\varphi _{-\frac{1}{2}}(\mathbf{r}) & = & \frac{R_{0}(r)}{\sqrt{\pi }}\left(|\ts -\frac{1}{2}\rangle +\frac{\sqrt{3}}{2}\theta e^{i\varphi }|\ts -\frac{3}{2}\rangle \right),\nonumber \\
\varphi _{-\frac{3}{2}}(\mathbf{r}) & = & R_{0}(r)\sqrt{\frac{3}{4\pi }}\theta e^{-i\varphi }|\ts -\frac{1}{2}\rangle .\nonumber 
\end{eqnarray}
Here the four unit columns $|Jm_{J}\rangle =|\pm 3/2\rangle $, $|\pm 1/2\rangle $
form the basis in the pseudo-spin $3/2$ space, and we have left only
terms up to first order in the small polar angle $\theta $ (cf. Eq.
(\ref{plm})).

Proceeding now to the calculation of exchange interaction, we construct
asymptotically correct basis of the direct product of the ground states
on two centers in the form \begin{equation}
\psi _{m_{F}^{\mathrm{A}}m_{F}^{\mathrm{B}}}^{\alpha \beta }(\mathbf{r}_{1},\mathbf{r}_{2})=\chi ^{\alpha \gamma ,\beta \delta }(\mathbf{r}_{1},\mathbf{r}_{2})\varphi _{m_{F}^{\mathrm{A}}}^{\gamma }(\mathbf{r}_{1\mathrm{A}})\varphi _{m_{F}^{\mathrm{B}}}^{\delta }(\mathbf{r}_{2\mathrm{B}}).\end{equation}
The total two-electron wave function $\psi $ is again sought in the
form (\ref{expan}), where the condition $c_{m_{F}^{\mathrm{A}}m_{F}^{\mathrm{B}}}=-c_{m_{F}^{\mathrm{A}}m_{F}^{\mathrm{B}}}^{(P)}$
is required for the wave function $\psi $ to be antisymmetric in
particle permutation. Equation on $\chi $ is obtained by substitution
of this form into the two-hole Hamiltonian:\begin{eqnarray}
\alpha _{l}(D_{ij}^{\alpha \alpha '}\hat{\zeta }_{i}\partial _{1j}\chi ^{\alpha '\gamma ,\beta \delta }-D_{ij}^{\beta \beta '}\hat{\zeta }_{i}\partial _{2j}\chi ^{\alpha \gamma ,\beta '\delta }) & + & \nonumber \\
+\Bigl [-\frac{1}{R/2-\zeta _{1}}-\frac{1}{R/2+\zeta _{2}}+\frac{1}{R}+\frac{1}{r_{12}}\Bigr ]\chi ^{\alpha \gamma ,\beta \delta } & = & 0,\label{eq:20}
\end{eqnarray}
where we used the asymptotics $\partial _{i}\varphi \approx -\alpha _{l}\varphi \partial _{i}r=-\alpha _{l}\varphi \hat{\zeta }_{i}\sgn \zeta $
at large $\mathbf{r}=\hat{\f \zeta }|\zeta |$ (cf. (\ref{eq:50})).

We neglect the derivatives of $\chi $ over transverse coordinates
as they are, as before, of higher order in $1/R$ (the validity of
this approximation is verified by differentiating the resulting expression
for $\chi $). 

Moreover, in the spherical approximation we consider, the hole Hamiltonian
is invariant under transformations from the spherical point symmetry
group $K_{h}$ and not only from $O_{h}$. Therefore, in this approximation,
axis $\hat{\f \zeta }$ connecting the acceptor centers may be chosen
as the quantization axis of the moments $\mathbf{J}$, $\mathbf{L}$
and $\mathbf{F}$, and the problem acquires some resemblance with
the hydrogen problem studied in Section I. 

In particular, the equations (\ref{eq:20}) on $\chi $ decouple from
each other: \begin{eqnarray}
\alpha _{l}(D_{zz}^{\alpha \alpha '}\partial _{\zeta _{1}}\chi ^{\alpha '\gamma ,\beta \delta }-D_{zz}^{\beta \beta '}\partial _{\zeta _{2}}\chi ^{\alpha \gamma ,\beta '\delta }) & + & \nonumber \\
+\Bigl [-\frac{1}{R/2-\zeta _{1}}-\frac{1}{R/2+\zeta _{2}}+\frac{1}{R}+\frac{1}{r_{12}}\Bigr ]\chi ^{\alpha \gamma ,\beta \delta } & = & 0,\label{eq:20-1}
\end{eqnarray}
because matrices $D_{zz}^{\alpha \alpha '}$ and $D_{zz}^{\beta \beta '}$
are diagonal: \[
D_{zz}=\left(\begin{array}{cccc}
 1 & 0 & 0 & 0\\
 0 & m_{h}/m_{l} & 0 & 0\\
 0 & 0 & m_{h}/m_{l} & 0\\
 0 & 0 & 0 & 1\end{array}
\right),\]
as can be easily seen using the explicit form of the pseudo-spin matrix
$J_{z}$. Together with the boundary condition $\chi ^{\alpha \gamma ,\beta \delta }{}\to \delta ^{\alpha \gamma }\delta ^{\beta \delta }$
when either $\mathbf{r}_{1}\to -\mathbf{R}/2$ or $\mathbf{r}_{2}\to \mathbf{R}/2$,
this equation yields solution for $\chi ^{\alpha \gamma ,\beta \delta }$
diagonal in each pair of band indices. i.e. only the elements with
$\alpha =\gamma $, $\beta =\delta $ are non-zero.

Because of such diagonal structure, the one-hole functions in the
elements of the exchange integral matrix \begin{eqnarray}
 &  & J_{m_{F}^{\mathrm{A}}m_{F}^{\mathrm{B}}m_{F}^{\mathrm{A}}{}'m_{F}^{\mathrm{B}}{}'}=-2\alpha _{l}\sum _{\alpha \beta }\int _{\zeta _{1}=\zeta _{2}}\left(\chi ^{\alpha \alpha ,\beta \beta }\right)^{2}\label{eq:mFAmFB}\\
 &  & \times \varphi _{m_{F}^{\mathrm{A}}}^{\alpha }(\mathbf{r}_{1\mathrm{A}})\varphi _{m_{F}^{\mathrm{B}}}^{\beta }(\mathbf{r}_{2\mathrm{B}})\varphi _{m_{F}^{\mathrm{A}}{}'}^{\alpha }(\mathbf{r}_{2\mathrm{A}})\varphi _{m_{F}^{\mathrm{B}}{}'}^{\beta }(\mathbf{r}_{1\mathrm{B}})d\zeta d^{4}\f \rho _{1,2}.\label{eq:mAmB_V_mA'mB'}\nonumber 
\end{eqnarray}

At integrating wave functions (\ref{eq:phi}) over transverse coordinates
in (\ref{eq:mFAmFB}) each $\theta $ will produce an additional smallness
$R^{-1/2}$ as we have seen in Section I. In principal order we may
hence leave only functions $\varphi _{\pm \frac{1}{2}}(\mathbf{r})\approx R_{0}(r)|\ts \pm \frac{1}{2}\rangle /\sqrt{\pi }$
(each of them has also only one non-zero pseudospin component). 

The 16$\times $16 diagonal matrices $D_{zz}^{\alpha \alpha '}\delta ^{\beta \beta '}$
before $\partial _{\zeta _{1}}$ and $\delta ^{\alpha \alpha '}D_{zz}^{\beta \beta '}$
before $\partial _{\zeta _{2}}$ in (\ref{eq:20-1}) do not generally
coincide for $\alpha \ne \beta $, and hence equations for all the
elements of $\chi $ cannot be simply solved in terms of $\zeta _{12}$
as was done in atomic case. Nevertheless, to calculate (\ref{eq:mFAmFB})
in main order we only need $\varphi _{\pm \frac{1}{2}}$ as we explained
above. Equations (\ref{eq:20-1}) on $\chi $ for the components $\varphi _{\pm \frac{1}{2}}$
\emph{do} reduce to (\ref{eq:chieq}) with $\alpha \to \alpha _{l}m_{h}/m_{l}$.
We thus arrive to a complete analogy with the problem of exchange
between two hydrogenic centers in the ground state, i.e. degenerate
only over projection of spin $\frac{1}{2}$. In this analogy localized
wave functions are $R_{0}(r)/\sqrt{\pi }$ and the role of the projection
of ``spin'' $\frac{1}{2}$ is played by $m_{F}$ spanning the truncated
subspace $\pm \frac{1}{2}$.

We may, therefore, at once give the asymptotically correct terms:
the ground state will be the ``singlet'' \begin{equation}
\frac{1}{\sqrt{2}}(\psi +\hat{P}\psi )\frac{1}{\sqrt{2}}(\ts |\frac{1}{2},-\frac{1}{2}\rangle -|-\frac{1}{2},\frac{1}{2}\rangle )\label{eq:singlet}\end{equation}
 and the excited state will be the ``triplet''\begin{equation}
\frac{1}{\sqrt{2}}(\psi -\hat{P}\psi )\left\{ \begin{array}{l}
 |\frac{1}{2},\frac{1}{2}\rangle \\
 \frac{1}{\sqrt{2}}(|\frac{1}{2},-\frac{1}{2}\rangle +|-\frac{1}{2},\frac{1}{2}\rangle )\\
 |-\frac{1}{2},-\frac{1}{2}\rangle \end{array}
\right.,\label{eq:triplet}\end{equation}
where $\psi =\chi R_{0}(r_{1\mathrm{A}})R_{0}(r_{2\mathrm{B}})/\pi $
and $\chi $ is given by (\ref{chi}) with $\alpha =\alpha _{l}m_{h}/m_{l}$. 

Energies of the ``singlet'' and ``triplet'' states are $E_{0}\pm J(R)$
respectively, where for the exchange splitting we obtain

\begin{eqnarray}
J & = & -\frac{m_{l}}{m_{h}}\Gamma \left(\frac{1}{2\alpha }\right)\frac{2^{\frac{3}{\alpha }-2}R^{3-\frac{1}{2\alpha }}}{\alpha ^{2+\frac{1}{2\alpha }}}e^{-2\alpha R}\label{eq:48}\nonumber \\
 & \times  & \int _{0}^{1}\frac{P^{2}\left[\frac{R}{2}(1+t)\right]P^{2}\left[\frac{R}{2}(1-t)\right]}{(1+t)^{\frac{3}{2\alpha }-2}(1-t)^{\frac{1}{2\alpha }-2}}e^{\frac{t-1}{\alpha }}dt,\label{eq:48-2}
\end{eqnarray}
where $P(r)$ is defined as $P(r)=R(r)e^{2\alpha R}$ with $R(r)$
from numerical solution. For two atoms, when $m_{l}=m_{h}$, $\alpha _{l}=\alpha $
and the preexponent $P(r)=Ar^{1/\alpha -1}/2$, Eq. (\ref{eq:48-2})
reduces to (\ref{Vij4}). 

Asymptotics $R_{0}(r)\approx R_{2}(r)=R(r)$ used to obtain $J(R)$
unfortunately sets in too far ($R>10$) and we have not evaluated
(\ref{eq:48-2}) numerically.

Similar to the hydrogen molecule problem the two states found, (\ref{eq:singlet})
and (\ref{eq:triplet}), with energies split by the amount of $2J(R)$
(\ref{eq:48-2}) represent only the ground and the most excited states.
Energies of all the other two-center states are defined by smaller
exchange integrals and will have exchange splitting $O(R^{-1}J(R))$,
i.e. at large $R$ these terms will lie between the first two. 

As for the magnetic moment of a single acceptor center, it is proportional
to $\mathbf{F}$ with the $g$-factor \cite{malyshev97} $g\approx 0.76$.
We see immediately that the ground state of two acceptor centers is
non-magnetic as is the case with hydrogenic atoms and donors.

\section{Role of magnetic field\label{sec:Role-of-magnetic}}

To complete the above discussion, we briefly review the impact of
external magnetic fields. Strong magnetic field squeezes electron
wave functions in cross direction reducing their overlap. Therefore,
major changes come already in exponential factors, and one may use
well-known results for the asymptotic behavior of hydrogenic wave
functions in magnetic fields.

Uniform magnetic field $\mathbf{H}=H\hat{\mathbf{z}}$ with the vector
potential $\mathbf{A}(\mathbf{r})=\frac{1}{2}\mathbf{H}\times \mathbf{r}$
results in an additional magnetic parabolic potential in the Schr\"odinger
equation for the ground state\begin{equation}
\left[-\frac{1}{2}\f \partial ^{2}-\frac{1}{r}+\frac{\rho ^{2}}{8}\left(\frac{a_{\mathrm{B}}}{a_{H}}\right)^{4}\right]\varphi =E_{H}\varphi ,\end{equation}
where $a_{H}^{2}=\hbar c/|e|H$ is the magnetic length. In weak magnetic
fields $a_{\mathrm{B}}\ll a_{H}$ magnetic potential is weak in the
region of localization $r\sim a_{\mathrm{B}}$, and $E_{H}\approx -\frac{1}{2}$.
High magnetic fields $a_{\mathrm{B}}\gg a_{H}$ localize an electron
in the transverse directions much stronger than Coulomb potential
does, and the binding energy grows logarithmically with magnetic field
$E_{H}=-\frac{1}{2}\ln ^{2}(a_{\mathrm{B}}/a_{H})^{2}$.

Exponential tails of localized electron wave function in magnetic
field have the following form (see \cite{shklovskii84} and references
therein). For a weak field ($a_{H}\gg a_{B}$) there are two asymptotic
regions with different dependencies of $\varphi (r)$. Rather close
to the $\hat{\mathbf{z}}$ axis, $1\ll r\ll (a_{H}/a_{B})^{2}$, wave
functions fall off as \begin{equation}
\varphi \propto \exp -r\left[1+\frac{\rho ^{2}}{24}\left(\frac{a_{\mathrm{B}}}{a_{H}}\right)^{4}\right],\end{equation}
whereas at large $r\gg (a_{H}/a_{B})^{2}$ as\begin{equation}
\varphi \propto \exp -\left[|z|+\frac{\rho ^{2}}{4}\left(\frac{a_{\mathrm{B}}}{a_{H}}\right)^{2}\right].\label{eq:strongFieldAsymptotics}\end{equation}
In strong magnetic fields ($a_{H}\ll a_{B}$) the first asymptotic
region is absent and exponential tail of the wave function is given
by \begin{equation}
\varphi \propto \exp -\left[|z|\ln \left(\frac{a_{\mathrm{B}}}{a_{H}}\right)^{2}+\frac{\rho ^{2}}{4}\left(\frac{a_{\mathrm{B}}}{a_{H}}\right)^{2}\right].\end{equation}
The change in the ``longitudinal'' asymptotics (field along the direction
connecting two centers) compared to (\ref{eq:strongFieldAsymptotics})
is due to the change of the binding energy (and hence, of the localization
length) with magnetic field.

From these formulas we observe that application of magnetic field
parallel to the inter-center axis does not change the form of the
exponential dependence of exchange integrals. At the same time, field
perpendicular to the axis changes exponential asymptotics either to
slightly corrected atomic one\begin{equation}
e^{-2R(1+R^{2}(a_{B}/a_{H})^{2}/6)}\end{equation}
for $1\ll R\ll (a_{H}/a_{B}^{2})$ in weak fields, or to \begin{equation}
e^{-R^{2}(a_{B}/a_{H})^{2}}\end{equation}
for $R\gg (a_{H}/a_{B}^{2})$ in weak fields and for all $R\gg 1$
in strong fields.

We conclude that magnetic field drastically changes already the exponential
dependencies, and we thus will not study the behavior of the pre-exponential
factors.

\section{Summary\label{sec:Conclusions}}

We have calculated exact asymptotic form of exchange interaction between
shallow impurity centers in semiconductors at large separations generalizing
the methods of the theory of hydrogenic molecules. 

We found that the ground state of a pair of impurity centers is always
non-magnetic. In other words, the exchange interactions between centers
have the ``antiferromagnetic'' sign.

The essence of the method that we use is the construction of a secular
equation with the matrix comprising exchange integrals between all
the degenerate single-center states. The exchange integrals are expressed
as the probability current through the hyperplane in the coordinate
space of two electrons that separates two regions of their localization
on the centers. The exchange integrals are calculated with wave functions
at the hyperplane corrected for the Coulomb interaction between electrons
and alien ions in principal order in inverse inter-center separation
$R^{-1}$.

Exchange integrals between different degenerate single-center states
may have different order in $R^{-1}$. The main exchange integrals
define the energy of the lowest and the highest split terms with the
remaining terms lying in between. To calculate the latter the exchange
integrals in the next approximation would be necessary. However, the
procedure simplifies for atoms if one uses symmetry considerations
to restrict the basis of degenerate states on each center to only
those that are allowed by symmetry to convert to a given term.

In semiconductors we have considered degeneracy of single-center states
both in valley index (donors) and in the band spectrum near $\mathbf{k}=0$
(acceptors). For electrons localized on donors calculation of exchange
interaction reduces to evaluation of exchange integrals between the
components of the wave functions from different pairs of valleys.
Depending on the relative position of the two donors in the cubic
lattice, crossover from one pair of valleys to another takes place
at high-symmetry directions. The dependence of exchange integrals
on valley parameters and on the orientation of the valleys with respect
to the axis connecting the two centers was calculated numerically.

We have also evaluated the asymptotic form of the asymmetric (Dzyaloshinskii-Moriya)
exchange interaction for donors.

The problem of exchange interaction between two acceptors simplifies
considerably in the spherical approximation when it acquires resemblance
with hydrogenic molecules. We have developed the complete procedure
for calculation of exchange in spherical approximation, but have restricted
ourselves to the sign of the interaction only. As in the case of a
hydrogen molecule, the ground state of a pair of acceptors turned
out to be non-magnetic. For a detailed calculation of the exchange
interaction numerical solution of the wave equation of an acceptor
center at very large distances would be required, which was not possible
to do in the numerical scheme we used.

The influence of external magnetic field on the strength of exchange
interaction was also briefly discussed. Expectedly, magnetic field
affects already the exponential dependence due to stronger localization
of electron on a center in the presence of magnetic field.

\section{Acknowledgments\label{sec:Acknowledgments}}

We are thankful to Dr. Sh. Kogan for referring us to the computational
method used to find asymptotics of wave functions numerically, to
Dr. B. L. Gel'mont for useful discussions, and to Dr. K. V. Kavokin
for referring us to \cite{malyshev97}. The work was supported (LPG)
by the NHMFL through the NSF cooperative agreement DMR-9527035 and
the State of Florida, and (PLK) by DARPA through the Naval Research
Laboratory Grant No. N00173-00-1-6005. 

\appendix

\section{Hierarchic structure of molecular terms\label{sec:Hierarchic-structure-of}}

In this Appendix we present asymptotic expressions for exchange integrals
between higher orbital momentum $l$ atomic states. Preexponents of
these integrals contain $R$ in different degrees dependent on the
projections $m$ of $l$ for the two atoms. We show how exchange integrals
(and hence electronic terms) of higher than leading order in $R^{-1}$
may be found by restricting the basis of single-atom wave functions
to only those allowed by symmetry to participate in a given molecular
term.

The matrix elements (\ref{Vij2}) for arbitrary $l$ were considered
in Ref. \cite{hadinger93} in the most general case of different binding
energies of the two atoms. However, these general formulas were cumbersome
and were not brought to explicit answer %
\footnote{Note that in Ref. \cite{hadinger93} due to incorrect change of variables
in (21) the overall multiplicative numerical coefficient is wrong.%
}. Below we cite our results for the case of equal binding energies:\begin{eqnarray}
J_{\nu _{\mathrm{A}}\nu _{\mathrm{B}}\nu '_{\mathrm{A}}\nu '_{\mathrm{B}}} & = & -\left(I_{\alpha +}+I_{\alpha -}\right)A_{\nu _{\mathrm{A}}}^{*}A_{\nu _{\mathrm{B}}}^{*}A_{\nu '_{\mathrm{A}}}A_{\nu '_{\mathrm{B}}}\nonumber \\
 & \times  & R^{\frac{7}{2\alpha }-1-\frac{m_{+}}{2}}e^{-2\alpha R},\label{Vij3}
\end{eqnarray}
 where \begin{eqnarray}
I_{\alpha \pm } & = & \frac{\Gamma \left(1/2\alpha \right)}{2^{3+\frac{1}{\alpha }-\frac{m_{+}}{2}}\alpha ^{2+\frac{1}{2\alpha }+\frac{m_{+}}{2}}}K(2\alpha ,m_{1},m_{2},\Delta m)\nonumber \\
 & \times  & \int _{0}^{1}(1-y)^{\frac{3}{2\alpha }\pm \frac{m_{-}}{2}}(1+y)^{\frac{1}{2\alpha }\mp \frac{m_{-}}{2}}e^{\frac{y-1}{\alpha }}dy\label{Iab+}
\end{eqnarray}
and \begin{eqnarray}
m_{+} & = & |m_{\mathrm{A}}|+|m'_{\mathrm{A}}|+|m_{\mathrm{B}}|+|m'_{\mathrm{B}}|\nonumber \\
m_{-} & = & |m_{\mathrm{A}}|+|m'_{\mathrm{A}}|-\left(|m_{\mathrm{B}}|+|m'_{\mathrm{B}}|\right),\nonumber \\
m_{1} & = & |m_{\mathrm{A}}|+|m'_{\mathrm{B}}|,\nonumber \\
m_{2} & = & |m'_{\mathrm{A}}|+|m_{\mathrm{B}}|,\nonumber \\
\Delta m & = & |m_{\mathrm{A}}-m'_{\mathrm{B}}|.
\end{eqnarray}

The four-fold integral \begin{eqnarray}
 & K & (2\alpha ,m_{1},m_{2},\Delta m)=\frac{2\alpha }{\pi ^{2}2^{1/2\alpha }\Gamma \left(1/2\alpha \right)}\int \rho _{12}^{1/\alpha }\rho _{1}^{m_{1}}\rho _{2}^{m_{2}}\nonumber \\
 &  & \times e^{-\rho _{1}^{2}-\rho _{2}^{2}}\cos \left[\Delta m(\varphi _{1}-\varphi _{2})\right]d^{2}\f \rho _{1}d^{2}\f \rho _{2}
\end{eqnarray}
may be calculated analytically. First three integrals are \begin{eqnarray}
K(2\alpha ,0,0,0) & = & 1,\nonumber \\
K(2\alpha ,2,0,0) & = & 1+1/4\alpha ,\\
K(2\alpha ,1,1,1) & = & -1/4\alpha .\nonumber 
\end{eqnarray}
 (See also \cite{hadinger93}). 

We observe from Eq. (\ref{Vij3}) that the matrix elements for transitions
between states with different $m$ have different degrees of $R$
in the preexponents. Since in all the calculations we have omitted
all but the leading terms in $R^{-1}$, secular equation (\ref{secular})
is valid only up to the accuracy of the matrix element with the highest
preexponent.

In the complete basis of all the degenerate atomic states the principal
matrix element in $R^{-1}$ is the one with minimal $m_{+}=0$, i.e.
between the states with $m_{\mathrm{A}}=m_{\mathrm{B}}=m'_{\mathrm{A}}=m'_{\mathrm{B}}=0$
(\ref{Vij4}). It corresponds to one of the molecular $\Sigma $ states.
With the accuracy up to this element, matrix $J_{m_{\mathrm{A}}m_{\mathrm{B}}m'_{\mathrm{A}}m'_{\mathrm{B}}}$
is trivial with only one finite element $J_{0}(R)$ at $m_{\mathrm{A}}=m_{\mathrm{B}}=m'_{\mathrm{A}}=m'_{\mathrm{B}}=0$.
So, at large $R$ molecular terms are arranged as follows: the ground
singlet $\Sigma $ state with the energy $E_{0}-|J_{0}(R)|$, the
triplet excited $\Sigma $ term $E_{0}+|J_{0}(R)|$, and the other
terms with energy is $E_{0}+O(R^{-1}J_{0}(R))$.

One can resolve the energies of those terms by restricting the basis
of degenerate atomic functions to only those that by symmetry can
convert to the molecular term with given $\Lambda =|m_{\mathrm{A}}+m_{\mathrm{B}}|=|m'_{\mathrm{A}}+m'_{\mathrm{B}}|$. 

\begin{figure}
\includegraphics[  width=0.99\columnwidth]{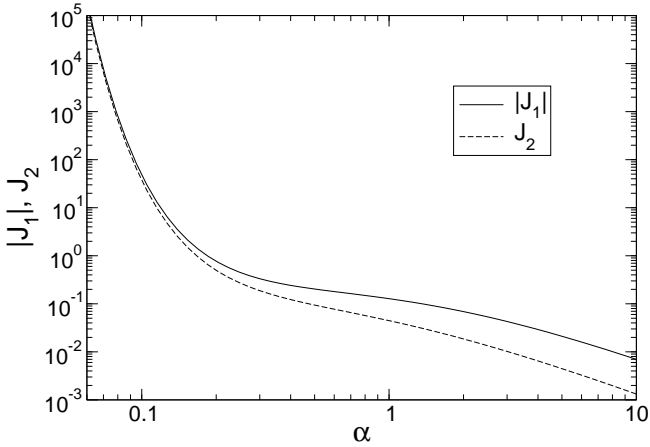}

\caption{Constants $J_{1}(\alpha )$ and $J_{2}(\alpha )$ from Eq. (\ref{eq:J}).\label{cap:Contants-J}}
\end{figure}

Because of the angular momentum projection conservation $m_{\mathrm{A}}+m_{\mathrm{B}}=m'_{\mathrm{A}}+m'_{\mathrm{B}}$
the next minimal $m_{+}=2$ will be for the matrix element between
the states with one of $m_{\mathrm{A}}$, $m_{\mathrm{B}}=1$, the
other zero. Such atomic states (and only them) convert to molecular
$\Pi $ states, which thus will have splitting $R^{7/2\alpha -2}e^{-2\alpha R}$.
Arranging the four states in the following order $|m_{\mathrm{A}}m_{\mathrm{B}}\rangle =|0,1\rangle $,
$|1,0\rangle $, $|0,-1\rangle $, $|-1,0\rangle $, we find that
the matrix $J_{m_{\mathrm{A}}m_{\mathrm{B}}m'_{\mathrm{A}}m'_{\mathrm{B}}}$
is\begin{eqnarray}
J & = & \frac{1}{4}|C_{l}|^{4}(2l+1)^{2}l(l+1)R^{7/2\alpha -2}e^{-2\alpha R}\nonumber \\
 & \times  & \left(\begin{array}{cccc}
 J_{2} & J_{1} & 0 & 0\\
 J_{1} & J_{2} & 0 & 0\\
 0 & 0 & J_{2} & J_{1}\\
 0 & 0 & J_{1} & J_{2}\end{array}
\right),\label{eq:matrixJ}
\end{eqnarray}
where \begin{eqnarray}
J_{1} & = & -\frac{1}{\alpha }\left(2+\frac{1}{2\alpha }\right)I(\alpha ),\nonumber \\
J_{2} & = & \frac{\Gamma \left(1/2\alpha \right)}{2^{3+\frac{1}{\alpha }}\alpha ^{4+\frac{1}{2\alpha }}}\nonumber \\
 &  & \int _{0}^{1}e^{\frac{y-1}{\alpha }}(1-y)^{\frac{3}{2\alpha }-1}(1+y)^{\frac{1}{2\alpha }-1}(1+y^{2})dy.\label{eq:J}
\end{eqnarray}
These constants are plotted as functions of $\alpha $ in Fig. \ref{cap:Contants-J}.

\begin{figure}
\includegraphics[  width=0.70\columnwidth]{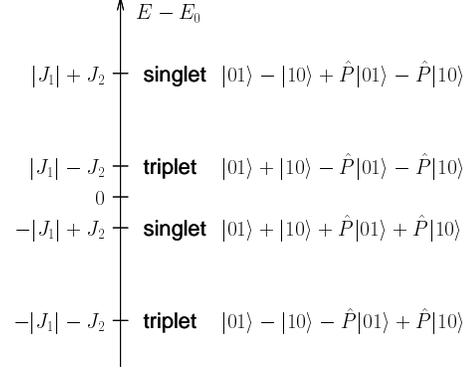}

\caption{Sketch of the energy level splitting of $p$-states of hydrogenic
atoms under exchange interaction in diatomic molecule (in units of
$\frac{1}{4}|C_{l}|^{4}(2l+1)^{2}l(l+1)R^{7/2\alpha -2}e^{-2\alpha R}$)\label{cap:Sketch-of-the}. }
\end{figure}

Matrix (\ref{eq:matrixJ}) has two doubly-degenerate eigenvalues $J_{2}\pm J_{1}\equiv J_{2}\mp |J_{1}|$.
Hence eight-fold degenerate two-electron state will split into four
doubly-degenerate coordinate states located in the order shown in
Fig. \ref{cap:Sketch-of-the}. The correct combination of two-electron
wave functions converting into molecular terms are presented in Fig.
\ref{cap:Sketch-of-the} for the total projection of the orbital momentum
$L=1$. The second degenerate state with $L=-1$ for each term is
given by substitution of 1 with $-1$ in the two-electron wave functions.

Note that unlike the energy level splitting of $s$-terms, the lowest
energy term of the two atoms each in a $p$-state turns out to be
spin-triplet while the most excited --- spin-singlet.

\end{document}